\documentclass[a4paper]{article}

\usepackage{url,hyperref,lineno,microtype,siunitx,subcaption,booktabs}
\usepackage[onehalfspacing]{setspace}

\usepackage{natbib,amsmath,booktabs,graphicx,authblk}
\usepackage[left=3cm, right=3cm, top=3cm, bottom=3cm]{geometry}

\usepackage[english]{babel}

\usepackage{todo}

\usepackage{listings}
\usepackage{color}
\usepackage[warn]{textcomp}

\usepackage{siunitx}
\DeclareSIUnit\molar{\mole\per\cubic\deci\metre}
\DeclareSIUnit\Molar{\textsc{m}}
\DeclareSIUnit\spike{spike}
\DeclareSIUnit\bits{bits}

\usepackage[labelfont=bf, font={footnotesize}]{caption}

\definecolor{comments}{RGB}{0,0,113}
\definecolor{red}{RGB}{160,0,0}
\definecolor{green}{RGB}{0,150,0}

\newcommand{\CodeSymbol}[1]{\textcolor{green}{#1}}
\newcommand\Authors{Christian R{\"o}ssert\,$^{1,*}$,
	Christian Pozzorini\,$^{2}$,
  Giuseppe Chindemi\,$^{1}$,
  Andrew P. Davison\,$^{3}$,
  Csaba Eroe\,$^{1}$,
  James King\,$^{1}$,
  Taylor H. Newton\,$^{1}$,
  Max Nolte\,$^{1}$,
  Srikanth Ramaswamy\,$^{1}$,
  Michael W. Reimann\,$^{1}$,
  Willem Wybo$^{1,2}$,
  Marc-Oliver Gewaltig\,$^{1}$,
  Wulfram Gerstner$^{2}$,
  Henry Markram\,$^{1,4}$,
  Idan Segev\,$^{5,6}$,
  Eilif Muller\,$^{1,*}$}

\newcommand\Address{
$^{1}$Blue Brain Project, \'Ecole Polytechnique F\'ed\'erale de Lausanne (EPFL)
Biotech Campus, Geneva, Switzerland \\
$^{2}$Laboratory of Computational Neuroscience (LCN), Brain Mind Institute, \'Ecole Polytechnique F\'ed\'erale de Lausanne, Lausanne, Switzerland \\
$^{3}$Neuroinformatics group, Unit\'e de Neurosciences, Information et Complexit\'e, Centre National de la Recherche Scientifique, Gif sur Yvette, France \\
$^{4}$Laboratory of Neural Microcircuitry, Brain Mind Institute, \'Ecole Polytechnique F\'ed\'erale de Lausanne, Lausanne, Switzerland \\
$^{5}$Department of Neurobiology, Alexander Silberman Institute of Life Sciences, The Hebrew University of Jerusalem, Jerusalem, Israel \\
$^{6}$The Edmond and Lily Safra Centre for Brain Sciences, The Hebrew University of Jerusalem, Jerusalem, Israel
}

\newcommand{\Vonda}{V_{S}^{D}}
\newcommand{\Vonsa}{V_{S}^{S}}
\newcommand{\Voffa}{V_{S}}

\newcommand{\Vond}{V^{D}}
\newcommand{\Vons}{V^{S}}
\newcommand{\Iin}{i_{\text{syn}}}
\newcommand{\Vonsp}{V^{S\prime}}

\newcommand{\fVond}{V^{D}}
\newcommand{\fVons}{V^{S}}
\newcommand{\fIin}{i_{\text{syn}}}
\newcommand{\fVonsp}{V^{S\prime}}

\begin{document}
\onecolumn

\title{Automated point-neuron simplification of data-driven microcircuit models}

\author{\Authors} 

\affil{\Address}

\maketitle

\begin{abstract}

  A method is presented for the reduction of morphologically detailed microcircuit models to a point-neuron representation without human intervention.
  The simplification occurs in a modular workflow, in the neighborhood of a user specified network activity state for the reference model, the ``operating point''.
  First, synapses are moved to the soma, correcting for dendritic filtering by low-pass filtering the delivered synaptic current.
  Filter parameters are computed numerically and independently for inhibitory and excitatory input using a Green's function approach.
  Next, point-neuron models for each neuron in the microcircuit are fit to their respective morphologically detailed counterparts.
  Here, generalized integrate-and-fire point neuron models are used, leveraging a recently published fitting toolbox.
  The fits are constrained by currents and voltages computed in the morphologically detailed partner neurons with soma corrected synapses at three depolarizations about the user specified operating point.
  The result is a simplified circuit which is well constrained by the reference circuit, and can be continuously updated as the latter iteratively integrates new data.
  The modularity of the approach makes it applicable also for other point-neuron and synapse models.

  The approach is demonstrated on a recently reported reconstruction of a neocortical microcircuit around an \emph{in vivo}-like working point.
  The resulting simplified network model is benchmarked to the reference morphologically detailed microcircuit model for a range of simulated network protocols.
  The simplified network is found to be slightly more sub-critical than the reference, with otherwise good agreement for both quantitative and qualitative validations.

\tiny

\end{abstract}

\section*{Comments on version 2}

The following changes have been made to the document since version 1:
The soma-synaptic filter fitting approach using PRAXIS has been replaced by a new method to directly extract the filters for each dendritic compartment using a Green's functions approach.
Methods Section 2.1, Results Section 3.1 and Figures 1, 2, 4, 5 have been updated to reflect these changes. Furthermore the Discussion has been updated to incorporate the new findings on reduction of post-synaptic filter variability.

\section{Introduction}

To understand the brain, it could be said that we must simultaneously appreciate its daunting complexity, and grasp its essential mechanisms.
While modern experimental neuroscience offers us a perpetually expanding view of the former, there remain major barriers to achieving the latter, an integrated and synthesized view of pertinent experimental facts for the functional principles of brain systems.
A first-draft integrated view of a piece of the neocortex as recently been reported by an international collaboration \citep{markram_reconstruction_2015}, resulting in precisely defined mathematical models accounting for much of the known cellular and synaptic anatomy and physiology of this part of the brain.
A method for synthesizing the resulting mathematical models to a minimal form is, thus formulated, an ill-posed problem.  Minimal for what purpose?

On the other hand, (spiking) point-neuron network models are widely used in theoretical neuroscience to describe various functions of the brain \citep{Thorpe2002, maass2004fading, eliasmith2012large, zenke2015diverse}.
They are easy to analyze and numerically light-weight, making them suitable for real-time and accelarated execution on GPU-based and neuromorphic platforms \citep{nageswaran2009configurable, fidjeland2010accelerated, bruderle2009establishing, Galluppi2010, Bruderle2011, yavuz_genn:_2016}.
However, these models for the most part have been generated using adhoc assumptions rather than constraining them to biological data making them disjunct from modern experimental neuroscience \citep{Eliasmith20141}.
To bridge the gap between these two areas of neuroscience, the goal of our simplification approach was to derive a point-neuron network from the experimentally constrained morphologically detailed network model in an automated, repeatable and quantitatively verifiable way. The result is a data-driven point-neuron network model which can be continuously updated as the data-driven reference continues to integrate the latest experimental data, without the need for human intervention.

While approaches exist to simplify the morphological complexity of biophysical neuron models by reducing dendritic arbors \citep{marasco_fast_2012, marasco_using_2013}, the phenomenological approach pursued here focuses on point neuron models as the specific target.  The primary difficulty hereby is that the synapses which excite or inhibit separate dendritic branches of the neuron have to be moved to the somatic compartment, in a way that accounts for the transformations of the synaptic responses due to the intermingling dendritic cable \citep{Rall1138}.

For our simplification procedure, we further employ the idea of the ``operating point'' a construct widely used in e.g. nonlinear control theory where the complex system is linearized at the point of interest \citep{slotine_applied_1991}.
In our case, the point of interest is the detailed network during \emph{in vivo}-like conditions and background activity.
We use this \emph{in vivo} network state to constrain synaptic correction factors and to extract parameters for the point neuron models using synaptic current and membrane potential data of individual neurons.

To benchmark the approach, we examine the simplified network by repeating a barrage of \emph{in silico} experiments used for the validation of the reference morphologically detailed network model \citep{markram_reconstruction_2015}.
Besides providing unprecedented validation of the point neuron network model, this approach is fundamentally important, as it tells us if and when a certain property is lost during the simplification pipeline.

The detailed network model further serves as a reference model for point neuron models and the automated simplification pipeline makes it easy to iteratively update the simplified model as new data is integrated into the detailed circuit.
Our approach will further allow us bridge from point neurons to population density and mean-field models in a quantitative manner, allowing dynamical systems and phase-plane analysis of the dynamics of data-driven networks.

To begin, we present a phenomenological method for soma-synaptic replacement and correction.
We show examples of this procedure for individual neurons and report on the quality of our approach.
Next, we show how detailed soma-synaptic corrected networks can be used to constrain point neuron models for each individual neuron in the network and we report on the results of constraining Generalized Integrate-and-Fire (GIF) models \citep{pozzorini_automated_2015}.
Finally we elaborate in detail on the validation of the simplified GIF point neuron network by comparing it to the detailed network based on synaptic physiology, synchronous-asynchronous spectrum of network states, sensory-evoked responses, temporal structure of in vivo spontaneous activity and spatial correlations.

\section{Methods}

\begin{figure}

\centerline{
\includegraphics[scale=1.0,keepaspectratio]{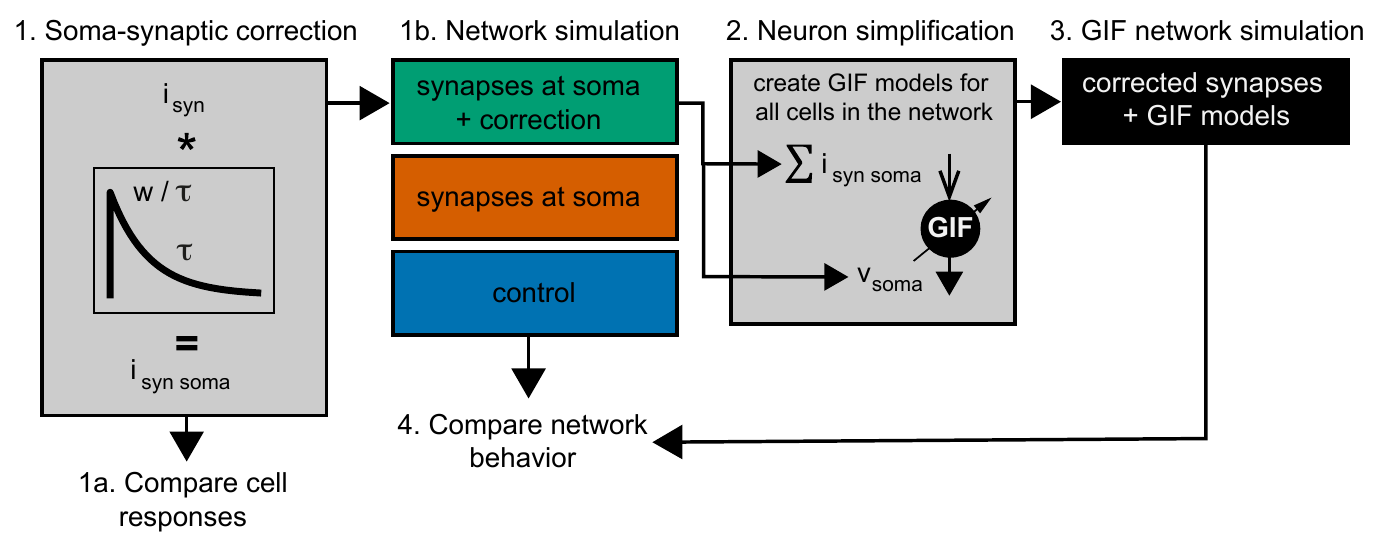}
}

 \caption{\textbf{Automated point-neuron simplification procedure.}
 The procedure of automated point-neuron simplification can be separated into two main parts:
 1. soma-synaptic correction and
 2. neuron simplification.
 }\label{fig:workflow}
\end{figure}

The automated point-neuron simplification procedure is modularized into two main steps: soma-synaptic correction and neuron simplification (Figure \ref{fig:workflow}).
In the ``soma-synatic correction'' step, all synapses on dendritic locations are moved to the soma while accounting for dendritic attenuation and delay (dendritic filtering) by applying a numerically computed corrective filter that approximately preserves the effective post-synaptic response (EPSP) of each synapse.
In the ``neuron simplification'' step, the total somatic stimulation- and synaptic currents and voltage responses for soma-synaptic corrected network simulations are used to constrain Generalized Integrate-and-Fire (GIF) models for each individual neuron in the network \citep{pozzorini_automated_2015}.

\subsection{Soma-synaptic correction}

\begin{figure}

\centerline{
\includegraphics[scale=1.0,keepaspectratio]{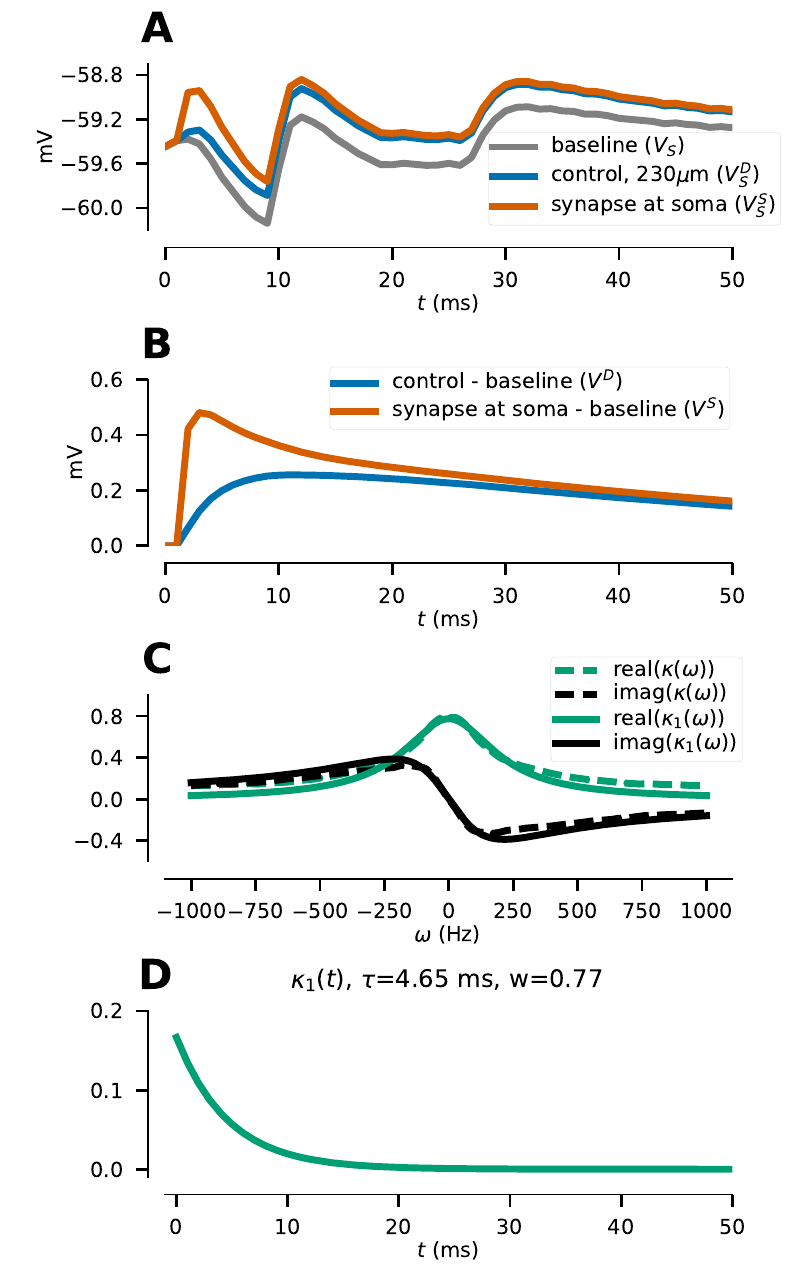}
}

 \caption{
 \textbf{Soma-synaptic correction method. A:}
 Resulting change in postsynaptic membrane potential recorded at the soma when moving a single synapse i from its dendritic location on the apical dendrite ($\Vonda(t)$, blue) at a L5 pyramidal cell (with path-distance of $230\,\mu \text{m}$ from the soma) to the soma ($\Vonsa(t)$, red). Baseline activity of the latter cell without activation of synapse i ($\Voffa(t)$, gray).
 \textbf{B:}
 Postsynaptic potentials of dendritically ($\Vond(t)$, blue) and somatically ($\Vons(t)$, red) activated synapse responses after subtraction of baseline activity.
 \textbf{C:}
 Frequency-domain response (real (green) and imaginary (black) parts) of full dendritic filter ($\kappa(\omega)$, dashed lines) extracted from baseline-subtracted dendritically and somatically activated postsynaptic responses and response of fitted single exponential filter ($\kappa_1(\omega)$, solid lines).
 \textbf{D:}
 Time-domain impulse response of single exponential filter ($\kappa_1(\omega)$, as extracted in C) with resulting parameters of $\tau=4.65\,\text{ms}$ and $w$=0.77.
 }\label{fig:syn_correction}
\end{figure}

To account for the dendritic filtering of synaptic responses which will be lost when moving them to the soma, all excitatory and inhibitory synapse models were extended by a simple low-pass filter of the synaptic current, $i_{\text{syn}}$, of the form
\begin{equation}
\label{eq_isyn_filt}
\tau \, \frac{d}{dt}i_{\text{syn soma}} = -i_{\text{syn soma}} + w \cdot i_{\text{syn}}
\end{equation}
where $w/\tau$ and $\tau$ are the amplitude and time constant of the filter respectively.
This filter approximates the effective attenuation and delay that a synaptic current would experience when flowing from its dendritic location to the soma \citep{Rall1138, berger2001high, williams2002dependence, nevian_properties_2007}.
The filter parameter estimation was however not conducted in the passive neuron, but during the replay of post synaptic activation coming from a network simulation mimicking physiological depolarization levels ($100\%$ rheobase) and extracellular calcium ($[Ca^{2+}]=1.25\,\text{mM}$) (Figure \ref{fig:syn_correction}A) \citep{markram_reconstruction_2015}.
Parametrizing the dendritic filtering during this bombardment of synaptic input, resulting in the so-called ``high-conductance state'' \citep{destexhe2003high}, approximates the neuron's operating point \emph{in vivo}, and therefore accounts also for first order effects of the interaction of dendritic non-linearities with on-going synaptic activity, and their role in altering dendritic properties such as input resistance, and shunting inhibition \citep{london_dendritic_2005}.

In the detailed network model synaptic connections between different neuron types have been grouped into different synaptic types being comprised of different synaptic properties, e.g. excitatory/inhibitory, ratio of NMDA/GABAB, synaptic decay time constants etc. to allow for variation in reported synaptic properties \citep{markram_reconstruction_2015}. The maximum number of different synaptic types is 12 with a distribution of $7.22 \pm 2.6$.

Ideally the parameters $w$ and $\tau$ have to be supplied for each individual synapse. However, to limit the computational cost of filter estimation, filter parameters for the same synaptic type and electrical compartment were only extracted once: for each synaptic type a ``probe'' synapse having respective mean synaptic parameters was either activated at each dendritic compartment ($\Vonda(t)$, Figure \ref{fig:syn_correction}A, blue trace) or at the soma ($\Vonsa(t)$, red trace).
In addition also the background activity without probe synapse stimulation was recorded ($\Voffa(t)$, gray trace).
After subtraction of the background activity

\begin{equation*}
\begin{aligned}
\Vond(t) = \Vonda(t) - \Voffa(t) \\
\Vons(t) = \Vonsa(t) - \Voffa(t).
\end{aligned}
\end{equation*}

the resulting traces (Figure \ref{fig:syn_correction}B) were centered around the time of spike arrival at the synapse of interest, so that the post-synaptic potential (PSP) starts at $t=0$. Our goal was to extract the parameters $w$ and $\tau$ \eqref{eq_isyn_filt} that modify the synaptic current so that the PSP when the probe synapse is at the soma optimally reproduces the PSP when the probe synapse is on the dendrites. To do so, we first derived a full kernel $\kappa$ in the Fourier domain that, when convolved with the synaptic input current when the probe synapse is at the soma, yields an exact reproduction of the PSP when the probe synapse is on the dendrites.
Then we approximated this kernel as a single exponential function
\begin{equation}\label{eq:expkernel}
\kappa_1(t) = \frac{w}{\tau} e^{-\frac{t}{\tau}} \approx \kappa(t).
\end{equation}
Convolving the synaptic current of the probe synapse with this exponential function is equivalent to \eqref{eq_isyn_filt}.

\paragraph{The kernel in the Fourier domain.} Following the Green's Function formalism \citep{Koch1998, Wybo2013, Wybo2015}, $\Vond(t)$ is found by convolving the synaptic input current $\Iin$ with a transfer kernel from dendrite to soma:
\begin{equation*}
\Vond(t) = \int_0^{\infty} \mathrm{d}s \, K_{D}(s) \, \Iin(t-s),
\end{equation*}
for which we will use the shorthand:
\begin{equation*}
\Vond(t) =  K_{D}(t) \ast \Iin(t).
\end{equation*}
Analogously, $\Vons(t)$ is found by convolving the synaptic input current with a somatic input kernel:
\begin{equation}\label{eq:Vonsoma}
\Vons(t) =  K_{S}(t) \ast \Iin(t).
\end{equation}

Our goal was to find a kernel $\kappa(t)$ that allows computing a modified somatic voltage when the synapse of interest is at the soma:
\begin{equation}\label{eq:VonsomaApprox}
\Vonsp(t) =  K_{S}(t) \ast \left( \kappa(t) \ast \Iin(t) \right),
\end{equation}
so that this voltage is as close as possible to the original voltage:
\begin{equation}\label{eq:goal}
\Vonsp(t) \approx \Vond(t).
\end{equation}

To do so, we work in the Fourier domain, where the convolutions become simple multiplications \citep{Bracewell1978}. Hence \eqref{eq:Vonsoma} becomes:
\begin{equation}\label{eq:fVonsoma}
\fVons(\omega) = K_{S}(\omega) \, \fIin(\omega),
\end{equation}
while for \eqref{eq:VonsomaApprox} we get:
\begin{equation*}
\fVonsp(\omega) = K_{S}(\omega) \, \kappa(\omega) \, \fIin(\omega).
\end{equation*}
Substituting \eqref{eq:fVonsoma} in this equation yields:
\begin{equation*}
\fVonsp(\omega) = \kappa(\omega) \, \fVons(\omega).
\end{equation*}
Converting \eqref{eq:goal} to the Fourier domain and combining it with the previous equation then allows the kernel $\kappa(\omega)$ to be computed from the Fourier transforms of the two known voltage traces:
\begin{equation}\label{eq:kappa}
\kappa(\omega) = \frac{\fVond(\omega) }{\fVons(\omega)}
\end{equation}

\paragraph{The kernel as a single exponential.} The Fourier transform of a single exponential function $c e^{at}$ ($a<0$) is given by the fraction $\frac{c}{i\omega + a}$.
Such a fraction was fitted to $\kappa(\omega)$ (Figure \ref{fig:syn_correction}C, dashed lines) as obtained from \eqref{eq:kappa} in the Fourier domain by means of the vector fitting algorithm \citep{Gustavsen1999, Wybo2015} resulting in $\kappa_1(\omega)$ (solid lines).
In equation \eqref{eq_isyn_filt}, $\tau$ is then given by $\tfrac{-1}{a}$ and $w$ is given by $\tfrac{-c}{a}$, resulting in the time domain representation $\kappa_1(t)$ (Figure \ref{fig:syn_correction}D)

\paragraph{Quality of filter correction} To test synaptic filter correction, $100$ excitatory neurons and inhibitory interneurons were randomly chosen from the microcircuit and were simulated for $40\,\text{s}$ with replay of identical network activity under three configurations:
\begin{enumerate}
\item control, with synapses at default locations;
\item synapses moved to the soma without correction and
\item synapses moved to the soma while applying soma-synaptic filter correction as described above.
\end{enumerate}
The quality has been estimated by comparing the coincidence factor $\Gamma$ \citep{jolivet_generalized_2004}, and the root mean square (RMS) voltage error of the non-spiking membrane potential response without sodium channels, mimicking bath application of tetrodotoxin (\emph{in silico} TTX), between configuration 1 and configuration 2 or 3.

\paragraph{Reduction of filter variability} While the filter weight can be easily incorporated into the synaptic weight of each synapse, the synaptic filter delay $\tau$ increases the complexity and variability of the synapses which counteracts the goal to increase simulation efficiency. We therefore also tested reduction of the variability of parameter $\tau$ by applying $k$-means clustering into sets of k = 1, 2, 3, 5 or 9 clusters using Lloyd's algorithm as implemented in the python function \textit{sklearn.cluster.KMeans}

\subsection{Neuron simplification}
\begin{figure}
\centerline{
\includegraphics[scale=1.0,keepaspectratio]{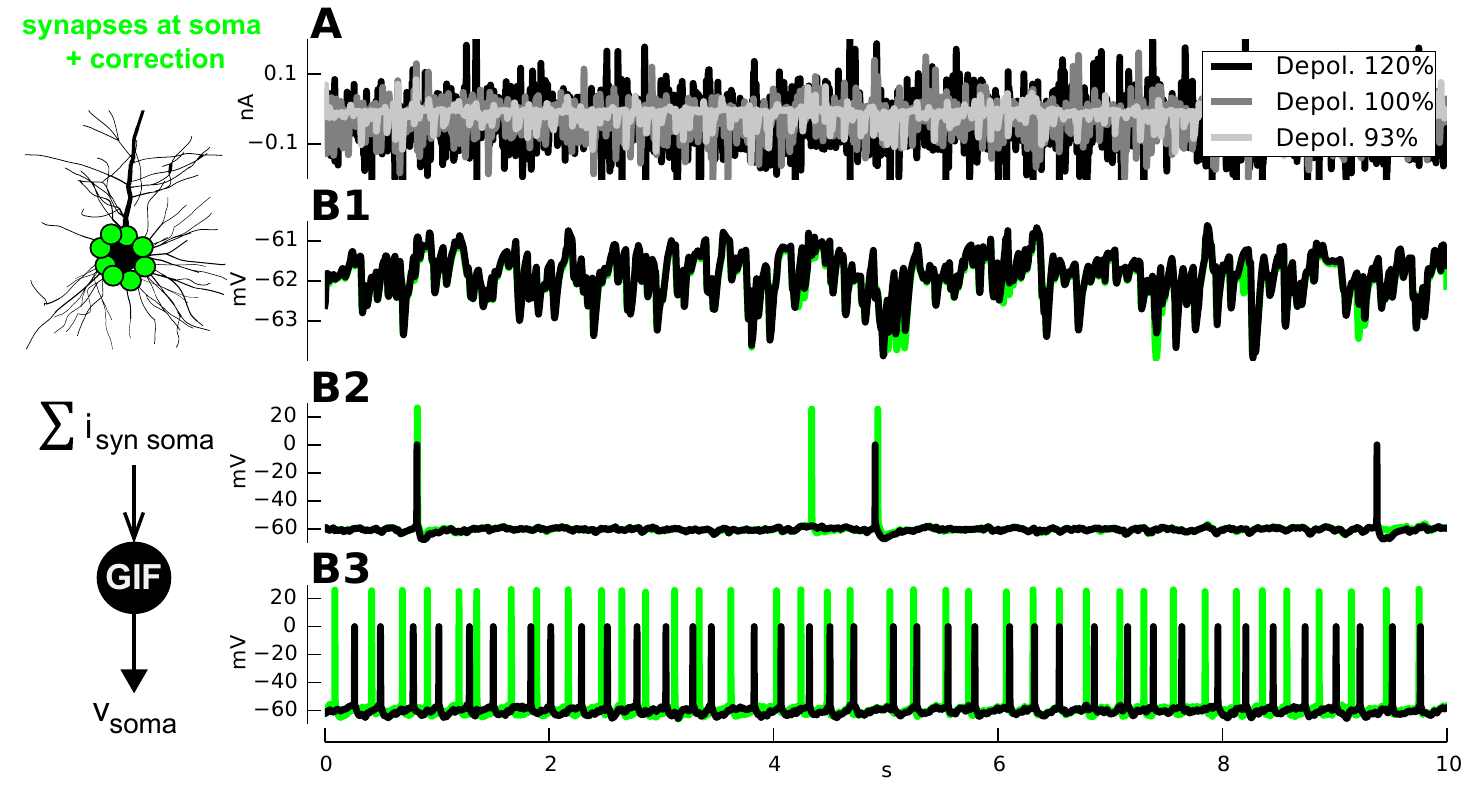}
}

 \caption{
 \textbf{GIF Model data generation and fitting procedure. A:}
 Recorded sum of synaptic currents during network simulation with depolarization levels (\% rheobase) of $93\%$, $100\%$ and $120\%$.
 \textbf{B:}
 Comparison of responses from detailed neuron (green) and GIF model (black) to corresponding depolarizing and synaptic currents (as shown in A) for $93\%$ (B1), $100\%$ (B2) and $120\%$ (B3).
 }\label{fig:gif_pipeline}
\end{figure}

Soma-synaptic corrected neurons were simplified to Generalized Integrate-and-Fire (GIF) point-neuron models about an \emph{in vivo}-like operating point using a previously published automated approach \citep{pozzorini_automated_2015}.
GIF model fitting was performed individually for each of $31346$ neurons in the microcircuit, and constrained to individual cell responses during network simulations mimicking physiological concentrations of extracellular calcium \emph{in vivo} ($1.25\,\text{mM}$) \citep{markram_reconstruction_2015}.
To broaden the generalization power of the GIF models, simulations for three different depolarization levels (expressed in \% rheobase of individual neurons) of $93\%$, $100\%$ and $120\%$ were used as constraints for the fitting process (Figure \ref{fig:gif_pipeline}).
For each of these three depolarization levels, four repetitions of $20\,\text{s}$ were simulated leading to a total length of $240\,\text{s}$, and divided into $180\,\text{s}$ used as a training set for GIF model parameter extraction, and $60\,\text{s}$ used as a validation set.
To generate current and voltage traces, network simulations were executed with all synapses moved to the soma and applying the synaptic correction (configuration 3).
This consequently allowed the recording of the total input current (due to depolarization and and synaptic inputs) at the soma for all depolarization levels (Figure \ref{fig:gif_pipeline}A), together with the individual membrane potential response of the cell (Figure \ref{fig:gif_pipeline}B, green traces), which served as constraints for the GIF model fitting process.

To allow for fast simulations, the spike-triggered current $\eta(t)$ [nA] and the spike-triggered firing threshold $\gamma(t)$ [mV] have been described by three exponential functions each.
For the spike-triggered current, the kernel $\eta'(t)$ extracted using rectangular basis functions \cite{pozzorini_automated_2015} was approximated by three exponentials.
Timescales of $\gamma(t)$ were fixed to $10$, $50$ and $250\,\text{ms}$ and only their amplitudes were optimized.
The refractory period was fixed to $T_{ref} = 3\,\text{ms}$ and $\lambda_0$ set to $1\,Hz$ leading to a total of $14$ extracted parameters per GIF model.
After the GIF parameter extraction procedure using the method described in \citep{pozzorini_automated_2015}, the performance of each GIF model was evaluated by estimating the log-likelihood $L$ i.e. probability that the validation spike train was produced by the model (\cite{pozzorini_automated_2015}, equation 20), and by comparing variance explained $\epsilon_V$ of membrane potential fluctuations between validation set and model (\cite{pozzorini_automated_2015}, equation 27).
Cells matching any of the following criteria
\begin{enumerate}
\item less than 5 spikes in the training set
\item Likelihood $< 2$ bits/spike
\item mean variance explained of membrane potential fluctuations of $\epsilon_V < 40\%$
\item $\Delta V >2$ mV
\end{enumerate}
were excluded from the results.
For the $6412$ cells which were excluded according to these criteria, they were replaced with a random GIF model drawn from successful optimizations from the same morpho-electrical class.

\subsection{Validations}

\subsubsection{Network simulations}
Simulations of a single cortical column containing a total of $31346$ neurons, as described in \cite{markram_reconstruction_2015} were conducted for four configurations as sketched above:
1. control,
2. synapses at soma,
3. synapses at soma + correction and
4. simplified GIF circuit (configuration 3. with all detailed neurons replaced by GIF neurons).

In brief, all simulations were executed on the EPFL Blue Brain IV BlueGene/Q supercomputer hosted at the Swiss National Supercomputing Centre in Lugano using NEURON \citep{hines1997neuron} with a time-step of $0.025\,\text{ms}$ unless otherwise specified.  Custom built software assisted in the model setup, definition of experiment, and online recording of state-variables of interest to disk, as described in \cite{markram_reconstruction_2015}.  Simulations results were analyzed in Python using standard scientific python tools and also custom software.

\subsubsection{Validation of synaptic physiology}

In the reference circuit, the kinetics and dynamics for individual synaptic contacts and connections were modeled by prescribing experimentally measured parameters, validated by comparing the amplitude, rise time, latency, decay time constant, transmission failure rate, and the coefficient of variation of resulting postsynaptic potentials (PSPs) against relevant \emph{in vitro} data, and corrected for extracellular $[Ca^{2+}]$ by downscaling unitary release probability, as described in \citep{markram_reconstruction_2015}.
In brief, excitatory synaptic transmission was modeled using both AMPA and NMDA receptor kinetics \citep{jahr_quantitative_1990, tsodyks_neural_1997, hausser_estimating_1997, markram_differential_1998, fuhrmann_coding_2002, ramaswamy_intrinsic_2012}.
Inhibitory synaptic transmission was modeled with a combination of GABAA and GABAB receptor kinetics \citep{khazipov_hippocampal_1995, de_koninck_endogenous_1997, mott_gabab-receptormediated_1999, gupta_organizing_2000}.
Stochastic synaptic transmission was implemented as a two-state Markov model of dynamic synaptic release by extending the Tsodyks-Markram dynamic synapse model to incorporate trial-to-trial variability \citep{tsodyks_neural_1997, fuhrmann_coding_2002}.

Apart from the inclusion of the additional low-pass filter given in (\ref{eq_isyn_filt}) to account for dendritic filtering as described above, synapse models used in the simplified microcircuit were otherwise unmodified from the reference.

To assess the correspondence of post-synaptic responses between the reference and simplified circuit, on average, about 100 pairs of pre-postsynaptic neurons at inter somatic distances of ~100 µm were chosen for each of the 1941 biologically viable pathways.
For individual pairs of pre-postsynaptic neurons at resting membrane potential, the presynaptic neuron was stimulated with a brief somatic current injection to elicit a unitary action potential, resulting in a postsynaptic response.
The postsynaptic response was determined as the average of 30 individual presynaptic stimulation trials.

\subsubsection{Sensory-evoked spike sequences}

The analysis of simulated single-whisker deflection \citep{reyes-puerta_laminar_2015} in the GIF neuron network (configuration 4) was performed on exactly the same cells as in the detailed reconstruction \cite{markram_reconstruction_2015}, with one exception as follows.
One L5 Martinotti cell was added as an exemplary OFF cell, since the previously chosen L5 Double Bouquet cell did not pass the significance test ($p > 0.05$, as used for classification by \cite{reyes-puerta_laminar_2015}) and is classified as not responding (NR) in the simplified circuit (it was OFF in the original circuit, the $p$ value changed slightly from below $0.05$ to above $0.05$).
All parameters are otherwise the same.
The firing rate is the average firing rate over $200$ trials from $-500$ to $500\,\text{ms}$, relative to the stimulus.

All cells in the circuit were included in the analysis with the exception of $21$ cells:
There are $18$ cells with a FR of more than $150$ Hz in the simplified circuit, and another $3$ cells with more than $100$ Hz in the original circuit. The former cells indicate an issue in the fallback solution for failed GIF model fits, which will be addressed in future versions of the automated simplification procedure.

\subsubsection{Temporal structure of in vivo spontaneous activity}

\cite{luczak_sequential_2007} report on the temporal structure of \emph{in vivo} spontaneous activity in the somatosensory cortex of anesthetized and awake rats.
They show that during periods of global activity (UP states), trios of neurons generate spike motifs with fixed temporal relationships that occur more frequently than predicted by chance.
Applying the same analysis techniques to the reference circuit yielded qualitatively similar results \cite{markram_reconstruction_2015}.
We sought to investigate whether neuron trios in our simplified circuit would also exhibit these properties.
Simulations were performed exactly as described in \cite{markram_reconstruction_2015} on the simplified circuit.
Briefly, to simulate UP state onsets, in the \emph{in vivo}-like state ($100\%$ depolarization, $[Ca^{2+}]=2\,\text{mM}$) we stimulated the center-most $20$ thalamic fibers of our circuit with single synchronous spikes after an initial $1500\,\text{ms}$ of relaxation time.
This experiment was repeated $25$ times, with each trial differing only in terms of random seeds.
From layer $5$ of the circuit, we selected $50$ cells (corresponding to 19600 unique cell trios) at random whose firing rate was greater than $3$ Hz over the $500\,\text{ms}$ active period.  Next, we concatenated the active period of each experiment, thus obtaining spike trains for each neuron with a duration of $12.5\,\text{s}$.
From the spike trains of the three cells in each trio, we calculated all possible spike triplets (see Figure \ref{fig:luczak}A), and extracted the mode (see Figure \ref{fig:luczak}B) for the count histogram of a representative neuron trio).
Count histograms were smoothed with a spatial Gaussian filter with a kernel of $10\,\text{ms}$.
Additionally, we created a scatter plot of individual neural latencies (defined as the average center of mass of the PSTH of a cell over a given UP state) against the inter-spike intervals associated with the triplet mode of the trios in which that cell participates.
Finally, we investigated two null hypotheses, namely, that spikes occur at random (the independent Poisson hypothesis) and that triplet structure is merely a result of increased overall activity shortly following UP state onset (the common excitability hypothesis).
To evaluate both hypotheses, the analysis was repeated with shuffled spike train data.
For the independent Poisson hypothesis, we preserved the number of spikes per train, but regenerated the individual spike times from a Poisson process.
For the common excitability model, we preserved the individual spike times and number of spikes per cell, but exchanged spikes at random between cells across time bins.
The procedures for both null hypotheses were repeated to obtain the mean and standard deviation over multiple trials.

\subsubsection{Spatial correlation}

The spatial correlation between local groups of neurons as function of distance between groups was investigated for the simplified circuit, and compared to the reference microcircuit model as described in \cite{markram_reconstruction_2015}.
In brief, the minicolumns of the microcircuit were grouped into local spatial clusters via a k-means clustering algorithm with a mean cluster size of $10$ minicolumns.
Mean PSTHs for each cluster were then computed using a time bin of $5\,\text{ms}$.
Pairwise cross-correlation coefficients were calculated between all clusters.
Distances between clusters refer to k-means centroid distances. Exponential fits were obtained using the python scipy.curve\_fit routine.

\subsubsection{Response reliability}

We compared the reliability of neuron responses between the reference and simplified circuits when stimulated by a single pulse stimulus delivered to $60$ thalamocortical fibers, as assessed by delay until the first spike after stimulation as described in \cite{markram_reconstruction_2015}.

\section{Results}

\subsection{Synaptic replacement and correction}

\begin{figure}
\centerline{
\includegraphics[scale=1.0,keepaspectratio]{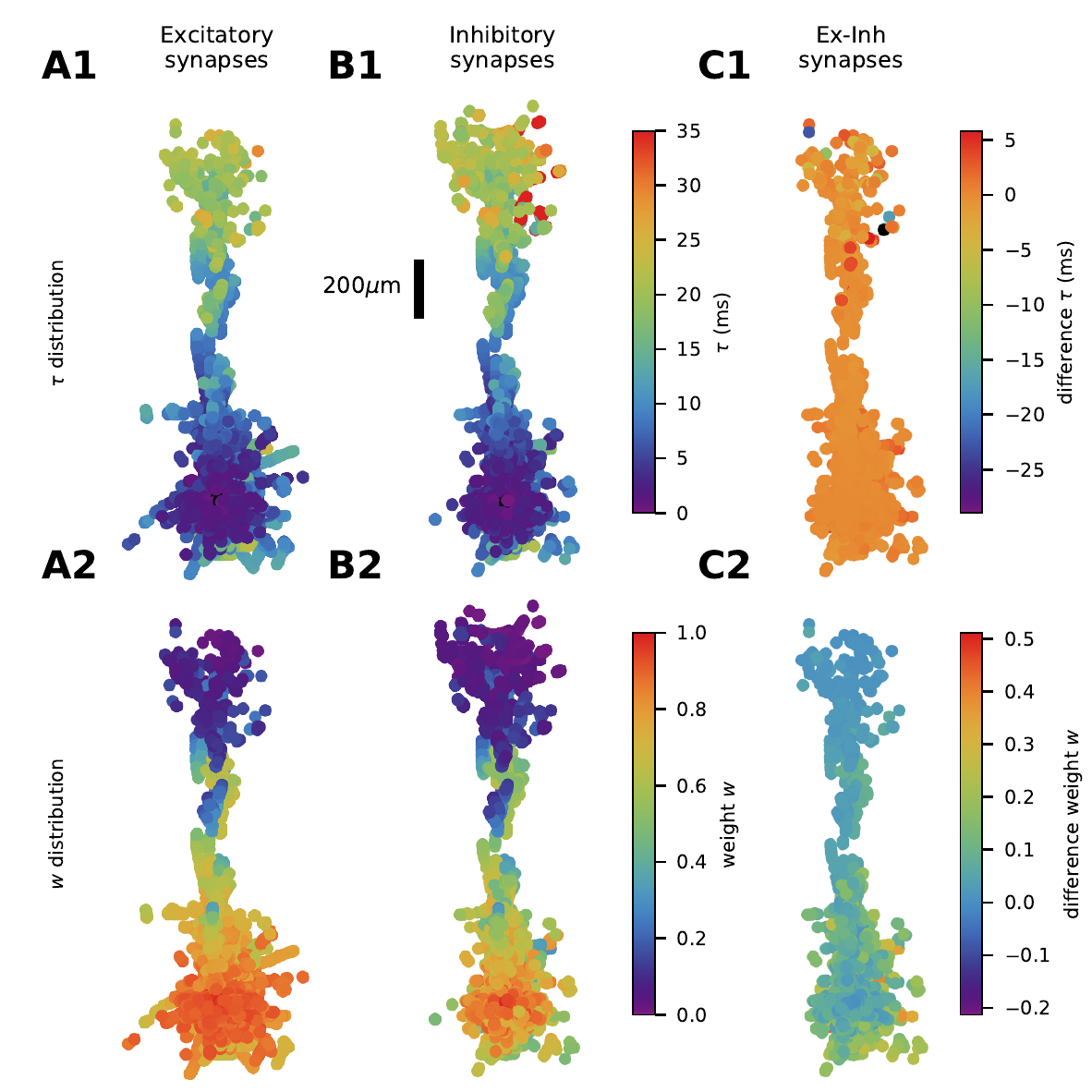}
}

 \caption{
 \textbf{Example of individual filters for Layer 5 thick-tufted pyramidal cells, type 1 (L5\_TTPC1). A,B:}
 Parameters of 10 L5\_TTPC1 neurons plotted as color coded points onto the xy plane of individual compartment locations.
 Location dependence for filter parameters $\tau$ (A1,B1) and $w$ (A2,B2) shown separately for excitatory (A) and inhibitory (B) synapses.
 \textbf{C:} Difference of filter parameters $\tau$ (C1) and $w$ (C2) between excitatory and inhibitory synapses for filters on the same compartment.
}
\label{fig:syn_factors}
\end{figure}

Synaptic correction filters properties $\tau$ and $w$ exhibited a systematic dependence on the distance between soma and the original dendritic location, as shown exemplarily for Layer 5 thick-tufted pyramidal cells, type 1 (L5\_TTPC1) (Figure \ref{fig:syn_factors}).
As expected the delay $\tau$ was lowest at the soma and reached values as high as $35\,\text{ms}$ (A1,B1) in the apical dendrites. Weight $w$ was high for filters close to the soma but synaptic signals had to be reduced by more than 90\% to account for signal attenuation in the apical dendrites (A2,B2).
While the difference for filter delay between excitatory and inhibitory synapses on the same compartments was insignificant ($\Delta \tau = -0.088 \pm 1.068\,\text{ms}$), the attenuation was higher in filters for inhibitory synapses on the same compartment than for excitatory synapses with $\Delta w = 0.062 \pm 0.054$. This justifies the need for separate filter properties per synapse type.

\begin{figure}

\centerline{
\includegraphics[scale=1.0,keepaspectratio]{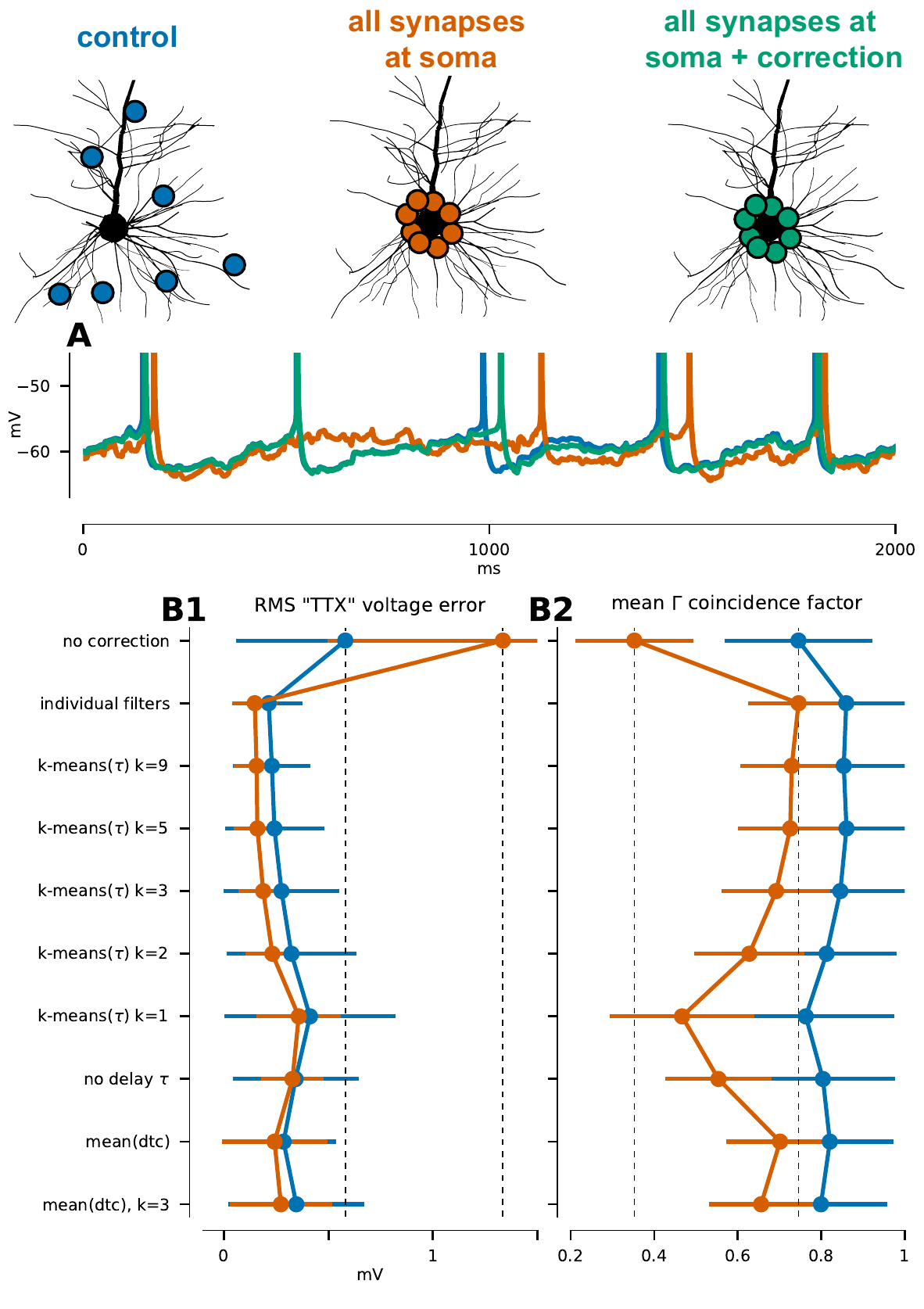}
}

 \caption{
 \textbf{Validation of synaptic filter correction. A:}
 Comparison of membrane potential responses for a L5\_TTPC1 cell during post-synaptic replay in the three configurations of control (green), synapses moved to the soma (red) and applied soma-synaptic correction (green).
 Action potentials in are cut.
 \textbf{B:}
 Comparison of mean and standard deviation of RMS membrane potential error (B1) and $\Gamma$ coincidence factor (B2) for 100 excitatory neurons (red) and 100 inhibitory interneurons (blue) for different soma-synaptic correction configurations. Results shown using no correction, using individual filters for each synapse, k-means binning of filter delays $\tau$, using no delay correction, using mean decay time constant for each synapse and in combination with k-means clustering (k=3).
 }\label{fig:syn_replaced_results}
\end{figure}

To validate our procedure of soma-synaptic correction, the different configurations of replaced uncorrected synapses (2) and replaced synapses + correction (3) were compared to the control case (1) during replay of synaptic activity in single neurons.
Direct examination of membrane potentials revealed that, especially for pyramidal cells, the inhibition was often too high and consequently the activity of these cells was reduced since inhibition was directly affecting the soma (Figure \ref{fig:syn_replaced_results}A, compare blue control and red uncorrected traces).
Application of the synaptic correction filter qualitatively restored the membrane and spiking dynamics (Figure \ref{fig:syn_replaced_results}A), compare blue to green traces).

Direct comparison of simulations without sodium currents (\emph{in silico} TTX) (Figure \ref{fig:syn_replaced_results}B1) showed that the error without soma-synaptic correction (``no correction'') was higher for excitatory (red) than for inhibitory interneurons (blue).
This error consequently lead to a mean spike coincidence smaller than 50\% in excitatory cells (Figure \ref{fig:syn_replaced_results}B2).
Applying the soma-synaptic correction for each compartment (``individual filters'') as estimated above greatly reduced the RMS error and increased the spike coincidence factor ($\Gamma$) especially for excitatory neurons (Figure \ref{fig:syn_replaced_results}B).

To further simplify the model and reduce synaptic variability we analyzed the grouping of filter delays $\tau$ using $k$-means clustering (Figure \ref{fig:syn_replaced_results}B, ``k-means($\tau$)''). This analysis revealed that only a small number of different filter delays ($k=3-5$) per synapse type were effective to obtain small RMS voltage errors and high coincidence factors. Surprisingly just using one cluster (``k-means($\tau$) k=1'') which is equivalent to applying the mean delay $\tau$ for all synapses was less effective than not using any filter delay correction at all (``no delay $\tau$'') which suggest that fast inputs close to the soma have been overcompensated in the former case.

To be able to significantly increase simulation speed also the variability of the synapses themselves have to be reduced. Here the main variability in the detailed model \citep{markram_reconstruction_2015} lies in a normal distributed decay time constant (dtc) of the post-synaptic process of the fast synaptic transmitters (AMPA or GABAA). Reducing this variability and using mean decay time constant for each of the synapse types (``mean(dtc)'') in combination with individual filters for each compartment increased RMS and decreased $\Gamma$. Combining this with three soma-synaptic filters (``mean(dtc), k=3'') slightly increases the error again while having the potential to greatly increase simulation speed by decreasing the number of post-synaptic processes to be simulated by two orders of magnitude (from thousands of synapses per cell to a maximum of $36 = 3 \cdot 12$ (maximum synaptic types)).

\subsection{Neuron simplification and validation}

\begin{figure}
\centerline{
\includegraphics[scale=1.0,keepaspectratio]{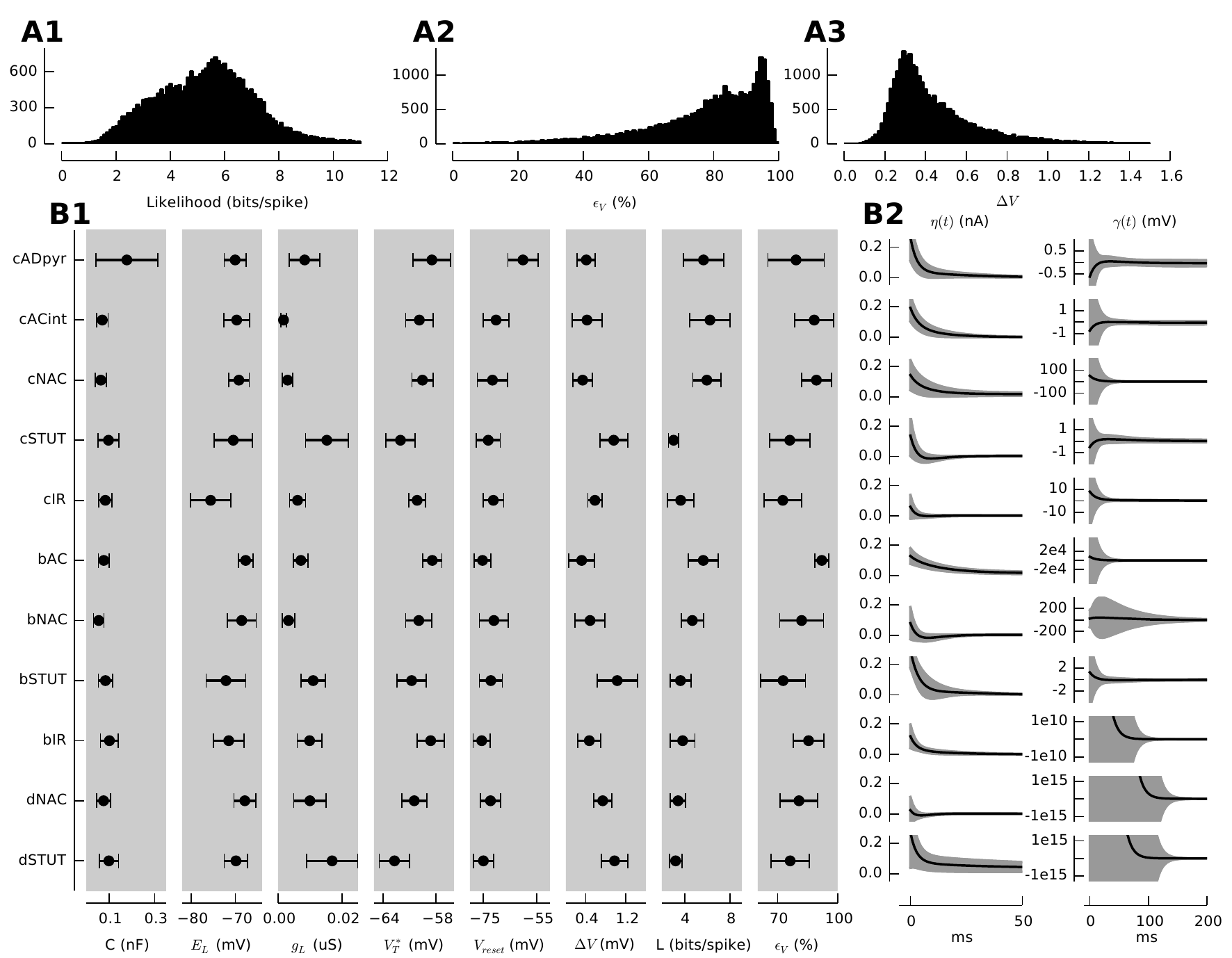}
}

 \caption{
 \textbf{Results of GIF Model fitting procedure. A:}
 Distribution of likelihood $L$ (bits/spike; see Methods) (A1), explained variance of the membrane potential fluctuations  $\epsilon_V$ (\%) (A2), and extracted level of stochasticity parameter $\Delta V$ (mV) (A3).
 \textbf{B1:}
 Mean and standard deviation for membrane capacity $C$ ($nF$), reversal potential $E_L$ (mV), membrane conductance $g_L$ ($\mu \text{S}$), threshold baseline $V_T^*$ (mV), voltage reset $V_{reset}$ (mV), and $\Delta V$, $L$ and $\epsilon_V$ grouped per electrical type of target neuron.
 \textbf{B2:}
 Mean and standard deviation of estimated filters for spike-triggered current $\eta(t)$ (left) and spike-triggered firing threshold $\gamma(t)$ (right) grouped per electrical type as indicated in A1.
 }\label{fig:gif_results}
\end{figure}

GIF model fitting was performed individually for each of the $31346$ neurons in the reference microcircuit, and constrained to individual cell responses during network simulations mimicking physiological concentrations of extracellular calcium \emph{in vivo} ($1.25\,\text{mM}$) \citep{markram_reconstruction_2015}.
In general GIF parameter extraction (Figure \ref{fig:gif_results}A) showed large likelihood (A1) and good explained variance of the membrane potential fluctuations (A2) for the majority of cells.
Since most of the optimized cells were deterministic, this can also be observed in the low extracted level of stochasticity $\Delta V$ (A3).
Only for the stochastic stuttering electrical types (cSTUT, bSTUD and dSTUT) was a higher $\Delta V$ observed (Figure \ref{fig:gif_results}B1).

In general the electrical cell types (e-types) showed a large variability in extracted parameters (Figure \ref{fig:gif_results}B1).
Consistent with our expectations, the extracted spike-triggered current $\eta(t)$ (Figure \ref{fig:gif_results}B2, left column) was most pronounced in adapting e-types (cADPyr, cACint, bAC) while it was fast in irregular (cIR, bIR) and most non-adapting types (bNAC, dNAC).
Interestingly some non-adapting e-types (cNAC) showed a large variability in $\eta(t)$. Furthermore, stuttering cells of type bSTUT and dSTUT showed longest time constants of spike-triggered currents.
For the spike-triggered firing threshold, all e-types showed a large variability in $\gamma(t)$, with bIR, dNAC and dSTUT types having the largest mean thresholds (Figure \ref{fig:gif_results}B2, right column).

\subsection{Synaptic physiology validation}

\begin{figure}

\centerline{
\includegraphics[scale=1.0,keepaspectratio]{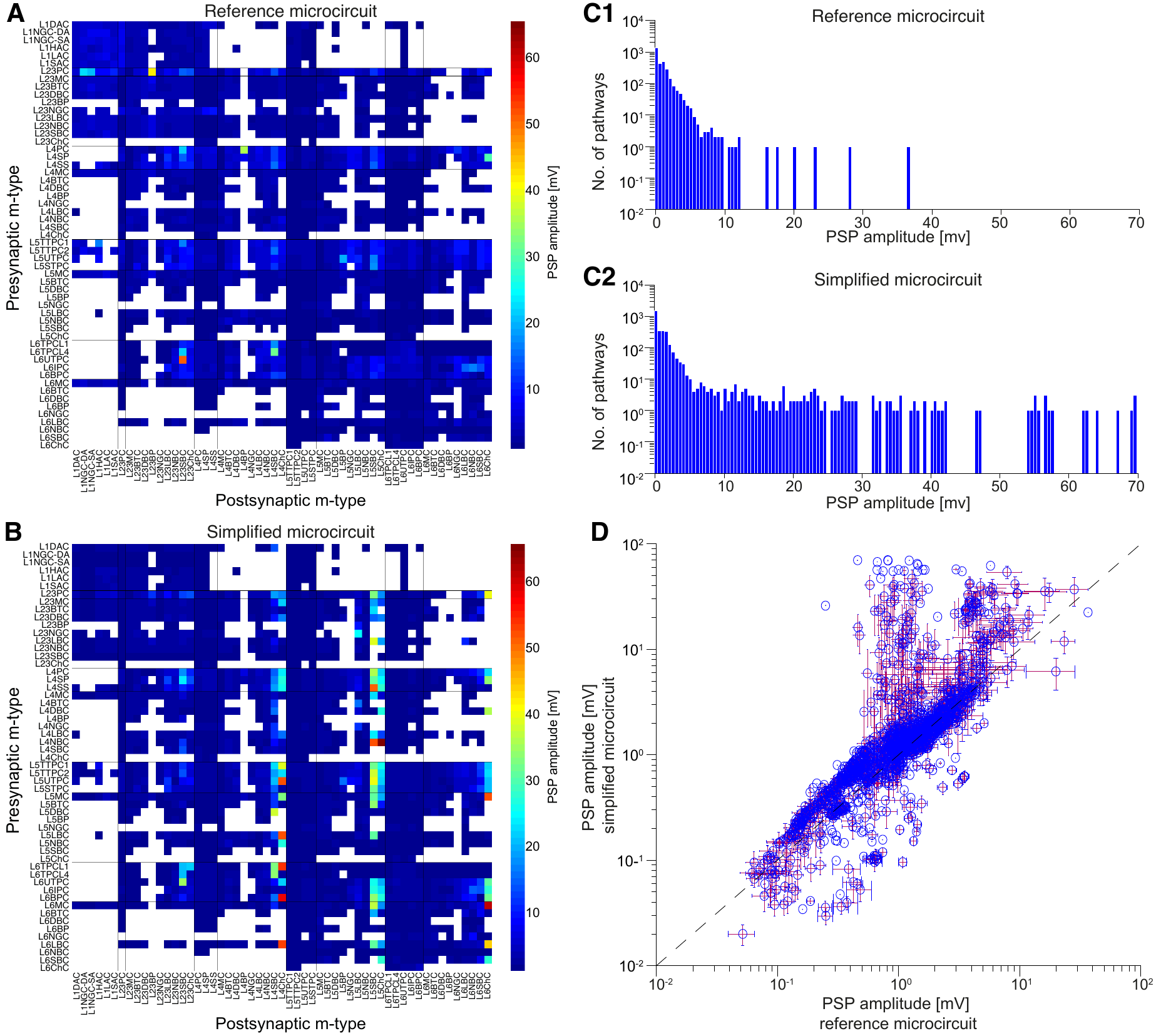}
}

 \caption{
 \textbf{Validating synaptic physiology. A:}
  Matrix representation of the average PSP amplitude for synaptic connections (pathways) formed between the 55 m-types (1,941 biological viable connections) in the reference microcircuit.
  \textbf{B:}
  Same as in A, but for the simplified microcircuit.
  \textbf{C1:}
  Distribution of PSP amplitudes. Histogram of the average PSP amplitude for 1,941 biologically viable connections in the reference microcircuit.
  \textbf{C2:}
  Same as in C1, but for the simplified microcircuit.
  \textbf{D:}
  Comparison of PSP amplitudes.
  Comparison of average PSP amplitudes between the reference and simplified microcircuits. Error bars show the standard deviation (SD).
  Dashed line shows the identity line.
 }\label{fig:synaptic_physiology}
\end{figure}

To validate the post-synaptic potential (PSP) responses of the simplified circuit conform to the reference model, we compared the average PSPs between pairs of neurons of specific pre- and post-synaptic m-types.
Compared to the reference model, shown in Figure \ref{fig:synaptic_physiology}A, the simplified model PSP amplitudes, shown in Figure \ref{fig:synaptic_physiology}B, are qualitatively similar.
Figure \ref{fig:synaptic_physiology}C reveals similar PSP amplitude distributions between the simplified and reference models, but with simplified circuit exhibiting a longer tail.
Figure \ref{fig:synaptic_physiology}D shows the numerical correspondence of average PSPs for the reference and simplified circuit.
While there is as a good degree of correspondence, some pathways are overly strong in the simplified circuit.

\subsection{Validation of network properties}

\subsubsection{Synchronous-asynchronous spectrum of network states}

\begin{figure}

\centerline{
\includegraphics[scale=1.0,keepaspectratio]{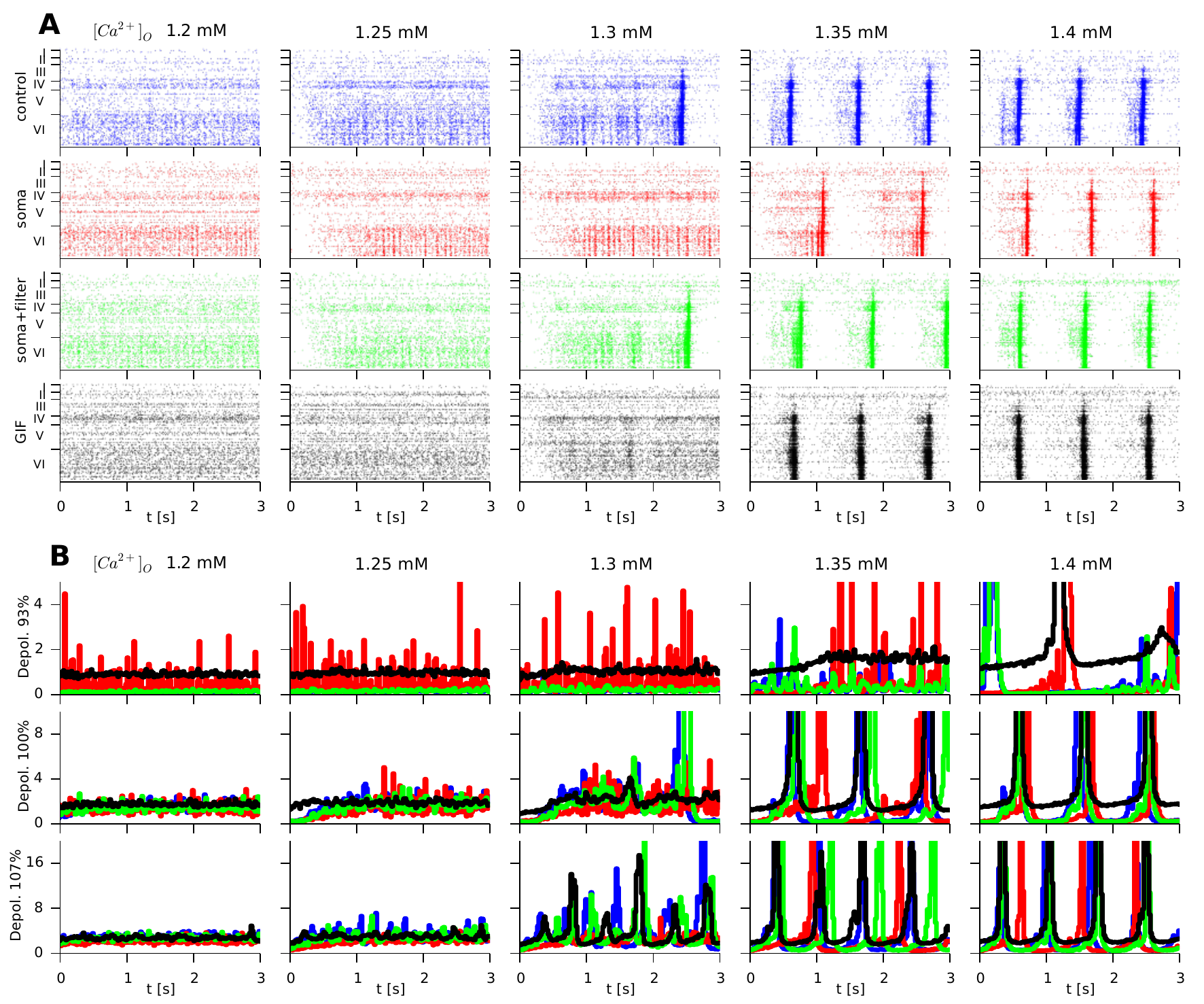}
}

 \caption{
 \textbf{Comparison of synchronous-asynchronous spectrum of network states. A:}
 Spike raster plots of the detailed control network (blue, first row), network with synapses moved to the soma (red, second row), network with synapses moved to the soma and soma-synaptic correction applied (green, third row), and network with all cells replaced by GIF models (black, fourth row). Exploration of emergent network states for a range of $[Ca^{2+}]_o$ from $1.2$ to $1.4\,\text{mM}$ at one depolarization level of 100\% threshold.
 \textbf{B:}
 Exploration of three different depolarization levels: $93$\% (first row), $100$\% (second row) and $107$\% (third row)) and range of different $[Ca^{2+}]_o$ concentrations (horizontal axis) as in A, assessed by peristimulus time histogram (PSTH).
 As before, comparison of four different configurations: control network (blue), all synapses replaced to soma (red), soma-synaptic correction applied (green) and GIF network (black).
 }\label{fig:raster_psth_networks}
\end{figure}

To evaluate the effect of the various stages of simplification on the basic network behavior, the microcircuit was simulated under various depolarization and calcium conditions (Figure \ref{fig:raster_psth_networks}).
In addition to the three previous configurations 1. control (blue), 2. synapses moved to the soma without correction (red) and synapses moved to the soma + correction (green) the fourth condition examined was the network with relocated synapses and GIF replacement of all cells (black).

The most prominent change seen from the raster plots with $100$\% threshold depolarization is that simply replacing the synapses to the soma leads to a shift in critical calcium concentration for network oscillations from $1.3$ to $1.35\,\text{mM}$ (compare first row (blue) to second row (red)).
Furthermore the initial frequency of oscillations was reduced but increases back to $1$ Hz like in the control case for $1.4\,\text{mM}$ calcium.
This change in emergent synchronous behavior can however be recovered completely when accounting for dendritic filtering by applying soma-synaptic correction (third row, green).
Even exchanging all neurons by the GIF equivalents retains the basic synchronous behavior (last row, black) as seen in the control.

Further investigating the cumulated spiking activity (PSTH) for different depolarization conditions ($93$, $100$ and $107$\%) (Figure \ref{fig:raster_psth_networks}B) reveals that without any synaptic corrections (red traces) the network shows prominent fast oscillations especially during low depolarization (first row) and lower calcium ($[Ca^{2+}]_o = 1.2$ to $1.35\,\text{mM}$).
These fast oscillations are however dominated by slow oscillations during high calcium ($1.4\,\text{mM}$).
Using soma-synaptic correction (green traces), a close match was achieved to the control configuration (blue traces) for all depolarization conditions leading to an almost exact match of bursting activity.
While this match in bursting activity was retained in the GIF neuron network (black traces), it did not generalize well to low levels of depolarization ($93$\%) leading to a higher baseline activity than control.
This difference can also be observed during oscillatory network states in all depolarization levels where the activity of the GIF network does not decay to zero between bursts.  These discrepancies highlight the need to pick an operating point for the fitting of point-neuron models to a morphologically detailed reference.
For the targeted operating point of the simplification at \emph{in vivo}-like conditions ($[Ca^{2+}]_o = 1.25\,\text{mM}$ and $100$\% depolarization), the GIF neuron network shows a good agreement of baseline network activity to the control case, however without an initial equilibratory transient as seen in the detailed network.

\subsubsection{Sensory-evoked responses}

\begin{figure}

\centerline{
\includegraphics[scale=0.28,keepaspectratio]{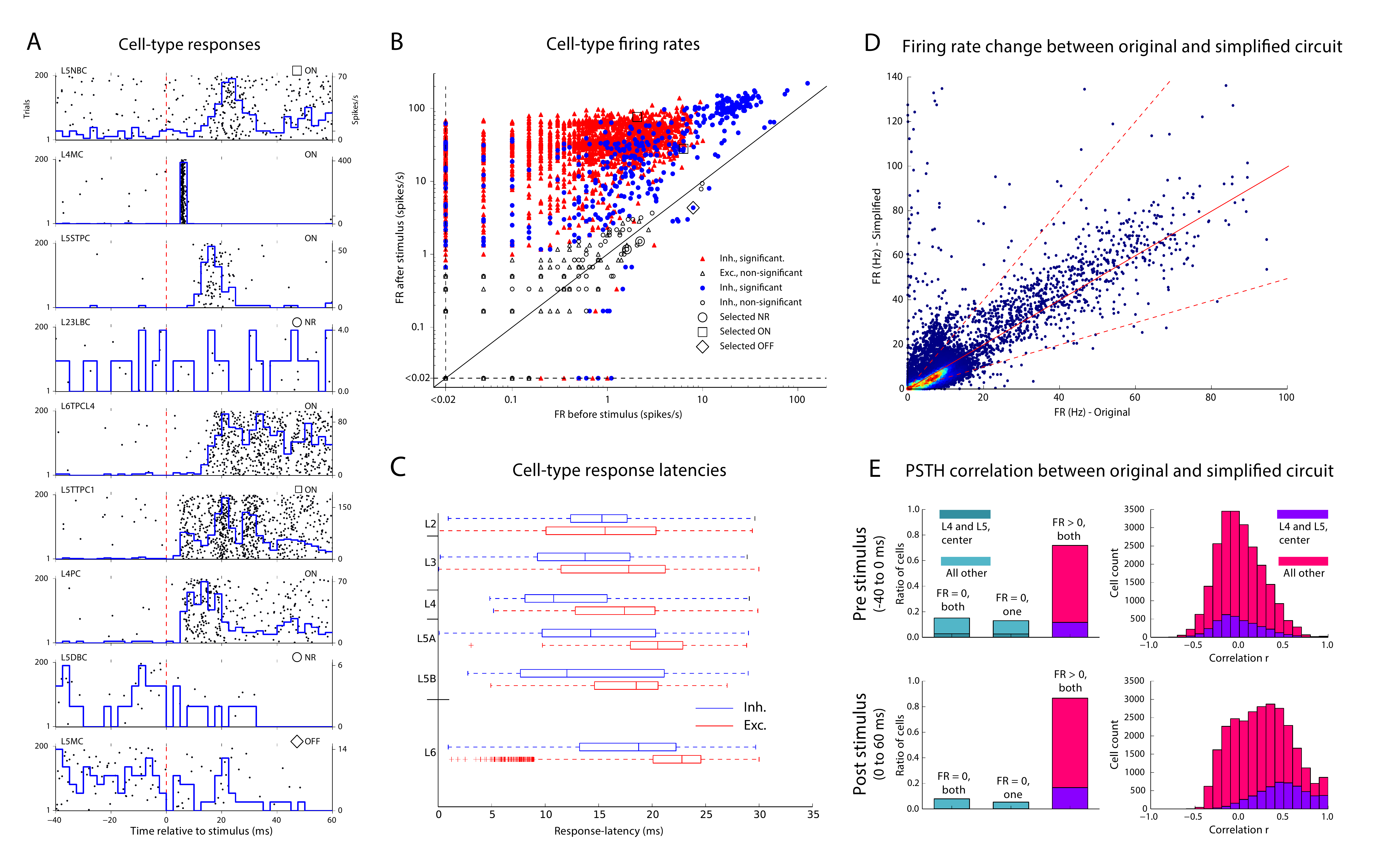}
}

 \caption{
 \textbf{Sensory-evoked spike sequences A:}
 Different cell response-types to simulated single-whisker deflection.
 The subplots depict the activity of the same cells as in the original detailed microcircuit experiment, showing PSTH and the aligned raster plot.
 After the stimulus, neurons either increased their firing rate (ON cells), showed no significant change in firing rate (NR cells), or decreased their firing rate (OFF cells).
 An additional OFF cell was added, since the depicted L5DBC is now classified as NR ($p > 0.05$).
 \textbf{B:}
 Comparison of mean firing rates before and after whisker deflection plotted in logarithmic scale ($2630$ excitatory and $550$ inhibitory neurons, same cells as in original microcircuit experiment).
 Empty symbols show NR cells, and filled symbols represent neurons showing significantly different ($p < 0.05$) activity (ON and OFF cells).
 \textbf{C:}
 Mean first-spike latencies of inhibitory (Inh.) and excitatory (Exc.) neurons to simulated whisker deflection (first spike occurrence within $30\,\text{ms}$ after stimulation, mean over $200$ trials, for all $31346$ neurons in the stimulated column)
 Each box plot represents median, interquartile, and range of latencies; crosses represent outliers ($2.5$ times interquartile range).
 \textbf{D:}
 Scatter plot showing the mean firing rate over $200$ trials of $1\,\text{ms}$ each ($500\,\text{ms}$ before stimulus to $500\,\text{ms}$ after stimulus), for almost all cells in the microcircuit.
 $3$ cells with a firing rate of over $100$ Hz in the original simulation are not included, and $18$ cells with a firing rate of over $140$ Hz in the simplified simulation are not included, since they are most likely due to failed GIF parameter extraction.
 Note that this is likely true for all neurons close to the simplified axis.
 Colors are according to Gaussian kernel density estimation (blue-low density, red-high density).
 The red line depicts a slope of $1$, and the dashed lines $0.5$ and $2$ respectively.
 \textbf{E:}
 Left: Ratio of cells that do not fire in either the original or simplified circuit ($40\,\text{ms}$ before, or $60\,\text{ms}$ after stimulus), do not fire in one of the simulation sets, or fire in both.
 Right: Linear correlation (Pearson's r) between PSTHs before and after the stimulus (as depicted in panel A).
 }\label{fig:reyes_puerta}
\end{figure}

To assess the stimulus response properties, several simulations from a study on the reference model where reproduced in the simplified model \cite{markram_reconstruction_2015}.
The PSTHs in response to simulated single-whisker deflection (see Methods) in the GIF neuron network, depicted in Figure \ref{fig:reyes_puerta}A, are qualitatively very similar to the results of the detailed circuit.
The PSTHs of the same eight cells as shown in \cite{markram_reconstruction_2015} are depicted, however, one L5 Martinotti cell was added as an exemplary OFF cell, since the chosen L5 Double Bouquet cell for the reference model no longer not passed the significance test ($p > 0.05$, as used for classification by \cite{reyes-puerta_laminar_2015}) in the simplified network.
A quantitative PSTH correlation analysis confirmed that there is a high positive linear correlation between the PSTHs in the detailed circuit (as computed in \cite{markram_reconstruction_2015} and the PSTHs in the GIF neuron network (see Figure \ref{fig:reyes_puerta}E).
The distribution of cell-type firing rates, in terms of firing rate before versus after the stimulus, is qualitatively similar too (see Figure \ref{fig:reyes_puerta}B), as are the first spike response latencies (Figure \ref{fig:reyes_puerta}C).
While overall mean firing rates are linearly correlated, some inhibitory cells have a sharply increased firing rate compared to the detailed microcircuit (see Figure \ref{fig:reyes_puerta}D), which indicates a fit-failure for certain cell types that will need to be fixed in future versions of the GIF model fitting module of the workflow.

\begin{figure}

\centerline{
\includegraphics[scale=1,keepaspectratio]{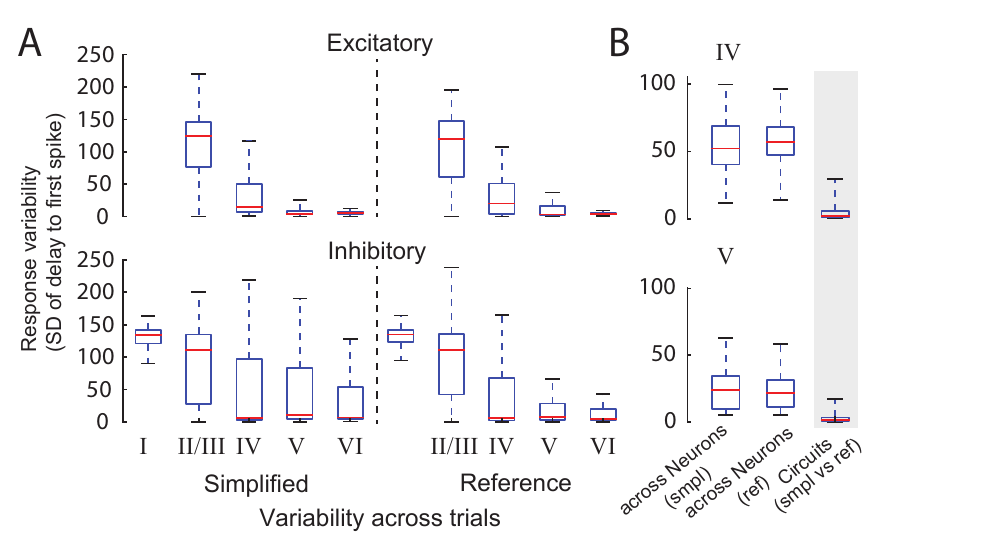}
}

 \caption{\textbf{Response variability. A:}
 Distributions of response variability, calculated as the SD over trials of the delay, for excitatory (top) and inhibitory (bottom) m-types.
 Red lines show the median, boxes the 25th and 75th percentiles and whiskers the full data spread (excluding outliers) across neurons.
 Left: In the simplified circuit, right: in the reference circuit.
 \textbf{B:}
 Distributions of response variability in layer IV (top) and layer V (bottom) compared to the variability between the reference and simplified circuit considering the same neuron in the same trial (grey shaded background).
 }\label{fig:variability}
\end{figure}

Response variability as assessed by the SD of the time-to-first-spike distribution across stimulation trials reveal good agreement between the reference and simplified models, as shown in Figure \ref{fig:variability}.

\begin{figure}

\centerline{
\includegraphics[scale=1,keepaspectratio]{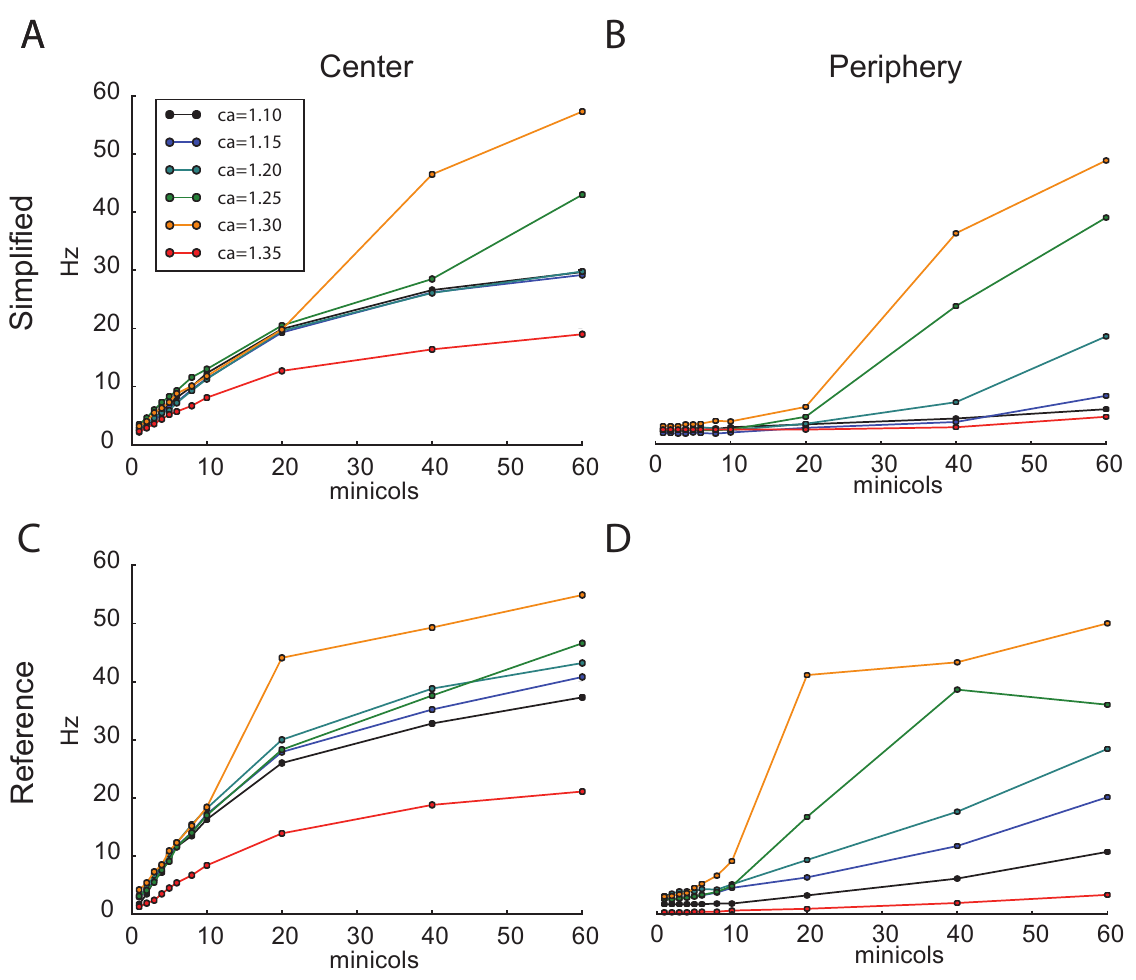}
}

 \caption{
 \textbf{Stimulus response curves for various levels of Ca2+ and somatic depolarization to 100\% threshold. A:}
 Response amplitudes defined as peak response of the $1000$ most central (A, C) or most peripheral (B, D) neurons in the $30\,\text{ms}$ after injection of thalamocortical stimuli of various strengths.
 (A, B) in the simplified circuit;
 (C, D) in the reference circuit.
 }\label{fig:response_curves}
\end{figure}

Next we assessed and compared stimulus response curves under \emph{in vivo}-like between the reference and simplified circuits.
Stimulating an increasing number of central thalamocortical fibers with a synchronous volley of spikes revealed initial central responses (max of PSTH within the $30\,\text{ms}$ after stimulus) that were qualitatively similar, but ~25\% reduced in the simplified circuit compared to the reference, over the range of stimuli and for $[Ca^{2+}]$ in the range $1.1-1.3\,\text{mM}$ (see Figures \ref{fig:response_curves}A and \ref{fig:response_curves}C).
Propagation of evoked responses to the periphery was markedly reduced in the simplified circuit.
As shown in Figures \ref{fig:response_curves}B and \ref{fig:response_curves}A, the peak response at the outermost neurons in the reference circuit showed a gradual increase with number of fibers stimulated (minicols) and increasing $[Ca^{2+}]$.
In contrast, the simplified circuit showed almost no dependence of the peripheral response on number of fibers stimulated (minicols) for $[Ca^{2+}]<1.25\,\text{mM}$.

\subsubsection{Temporal structure of \emph{in vivo} spontaneous activity}

\begin{figure}

\centerline{
\includegraphics[scale=0.5,keepaspectratio]{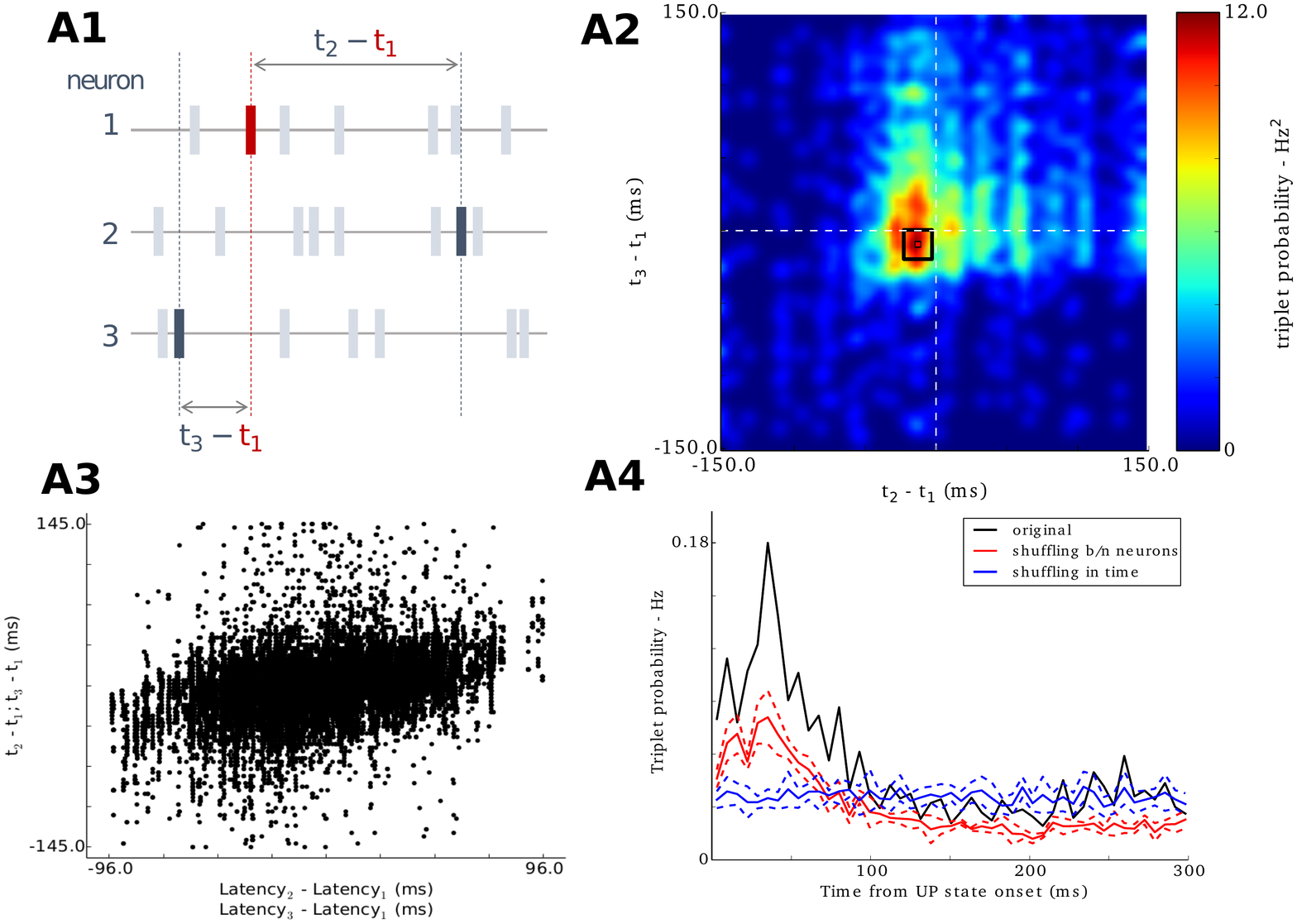}
}

 \caption{\textbf{Temporal structure of in vivo spontaneous activity. A:}
Schematic depiction of the structure of a spike triplet for a trio of neurons.  From a pool of 50 randomly selected L5 neurons, every possible triplet for every possible cell trio was computed to produce the data plotted in figures B through D.
\textbf{B}: Count histogram for a representative neuron trio.
Black box indicates region containing ``precisely repeating triplets'' (those that occur sufficiently close to mode).
White square signifies mode.
\textbf{C}: Plot of correlation between neural latency differences and inter spike intervals for corresponding triplet modes.
Positive slope indicates that individual neural latencies are predictive of triplet structure for trios in which those neurons participate.
Data points whose triplet inter spike intervals exceed $m \pm150\,\text{ms}$ are not shown.
\textbf{D}: Probability of occurrence of precisely repeating triplets peaks shortly after onset of activated state (black, solid line).
The significance of this peak is compared with two alternative null hypotheses (independent Poisson model, blue curve; common excitability model, red curve).
The independent Poisson model preserves overall number spikes per train, but draws spike times at random from a Poisson distribution.
The common excitability model preserves individual spike times, but randomly exchanges these spikes between neurons in the trio.
Dashed lines indicate standard deviation.
}\label{fig:luczak}
\end{figure}

As described in \citealp{markram_reconstruction_2015}, simulations of the reference model revealed temporal structure comparable to recent findings \emph{in vivo} \cite{luczak_sequential_2007}.
To assess if such temporal structure was preserved in the simplified network, the same analysis were performed for the simplified circuit.
As shown in Figure \ref{fig:luczak}, the observed precisely repeating triplets are also found to be present in the simplified circuit, albeit with significantly reduced frequency.
Increasing the extracellular calcium level to within $~0.02\,\text{mM}$ of the critical point for the simplified network markedly increased the frequency to exceed those levels found in the reference circuit, indicating disagreement between the simplified and reference circuit on this validation are due to the minor differences in level of criticality between the two.
This highlights the need to develop a calibration technique to make network quantitatively compatible under validations which are criticality-level dependent.

\subsubsection{Spatial correlation}

\begin{figure}

\centerline{
\includegraphics[scale=0.3,keepaspectratio]{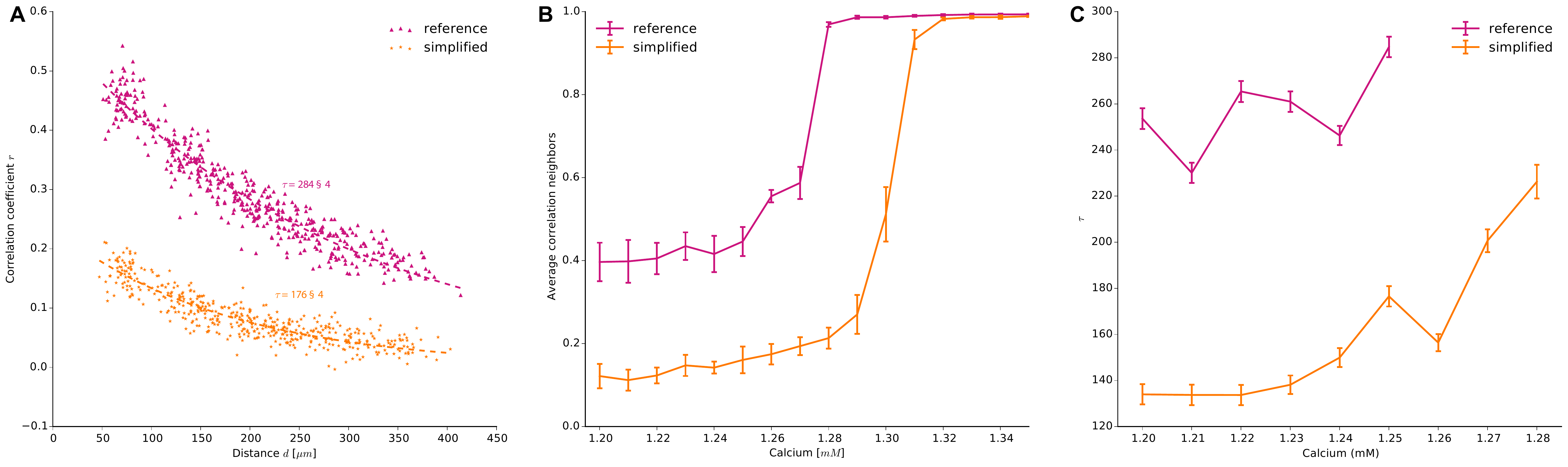}
}

 \caption{\textbf{Spatial correlation. A:}
 Correlation Scatter.
 The pairwise activity correlation coefficient between clusters of mini columns decays with the distance of the clusters.
 Overall, the correlation is lower in the simplified circuit.
 Results obtained at $1.25\,mM$ extra cellular calcium.
 Exponential fits showed by dashed lines.
 \textbf{B:}
 Neighbors Average Correlation.
 The average correlation between neighboring clusters with a maximum distance of $60\,\mu \text{m}$ increases with extra cellular calcium concentration, with a sharp jump at the transition point from the spontaneous activity regime to the
 synchronous one. Error bars show error of fit.
 \textbf{C:}
 Space Constant.
 Fitted space constant $\tau$ (see panel A) as a function of extra cellular calcium concentration.
 Higher calcium concentration is needed in the case of the simplified circuit to reach the same tau of the detailed circuit.
 For higher calcium concentrations, correlation data are no longer well approximated by an exponential curve (data not shown).
 Error bars show error of fit.
 }\label{fig:spatial_corr}
\end{figure}

The previously described spatial dependence of PSTH correlations between clusters of neurons in the reference circuit \citep{markram_reconstruction_2015}, is qualitatively replicated in the GIF neuron network (see Figure \ref{fig:spatial_corr}, left), but the simplified circuit has a significantly faster fall-off of spatial correlations ($\tau=176 \pm 4\,\mu \text{m}$ versus $284 \pm 4\,\mu \text{m}$ for the reference), and a significantly reduced correlation overall (see Figure \ref{fig:spatial_corr}, middle) for all calcium conditions except for in the immediate neighborhood of the critical point.
Plotting $\tau$ as a function of extracellular calcium concentration reveals (Figure \ref{fig:spatial_corr}, right) that only for calcium concentrations within $~0.02\,\text{mM}$ of the critical point, could the $\tau$ of the reference circuit at $1.25\,\text{mM}$ be approached.  Taken together, these simulations indicate that the simplified circuit appears to have genuinely reduced firing rate correlation between clusters of neurons compared to the reference circuit.

\section{Discussion}

Here we present a pipeline for automated point-neuron simplification of data-driven microcircuit models, and demonstrate the technique on a recently reported reconstruction of a neocortical microcircuit around an \emph{in vivo}-like working point.
Our modular method first uses a phenomenological approach that corrects for the effective PSP response when relocating synapses from the dendrites to the soma.
It then uses a high-throughput parameter extraction of Generalized Integrate-and-Fire (GIF) point-neuron models \citep{pozzorini_automated_2015} to replace each detailed neuron.
We consider first the findings and implications raised by our applied synaptic correction and cellular simplification and then discuss the results of network simulations and validations.

\subsection{Soma-synaptic correction and cellular simplification}

While many different approaches exist to simplify dendritic complexity by e.g. reducing the number of dendritic arbors, this approach is to our knowledge the first to target point-neuron models and directly and systematically correct for dendritic attenuation by modifying the synaptic dynamics directly.
Our approach is not intended and cannot account for non-linear local dendritic computation but is nevertheless very powerful in correcting for the change in synaptic delay and effectiveness when neurons have to be reduced to a point model.
It can reduce main errors in membrane activity of neurons especially with long dendritic arbors i.e. apical dendrites in pyramidal cells.
Soma-synaptic correction is most effective on the membrane potential error especially in simulations without non-linear activation of sodium channels (\emph{in silico} TTX) but also spiking precision can be recaptured to some extent (Figure \ref{fig:syn_replaced_results}B).

Using the soma-synaptic correction approach, it is further straight-forward to generate data that can be used to constrain simplified neuron models.
Since all synapses are located at the soma the sum of synaptic currents recorded during a simulation of the detailed network in combination with its somatic membrane response can be used to simplify the neuron around its operating point.
While this data can be used to constrain any point-neuron model, we chose the GIF neuron fitting procedure for its fast convergence properties.

Our extracted GIF models from these data showed overall good performance on the validation traces.
Nevertheless $20$\% of the cells had to be replaced by other GIF models of the same respective morpho-electrical type due to either non converging optimization or low performance on the validation set.

\subsection{Effect of simplification on network behavior}

Our soma-synaptic correction approach was able to recapture the transition from asynchronous to synchronous network states with increasing calcium concentrations as seen in the detailed control network simulations, with only minor shifts in the critical calcium concentration.
Furthermore it was able to remove fast oscillations especially appearing in layer 6 when synapses were displaced to the soma without compensation.
The basic asynchronous and synchronous behavior was retained when all detailed models were replaced by simplified GIF models (Figure \ref{fig:raster_psth_networks}).
However, GIF models could not recapture the low baseline activity as observed during low depolarizations or in between synchronous burst, where the cumulated spike activity did not decay down close to zero.

Stimulus response properties were largely preserved, though the simplified circuit was slightly less responsive.  Interestingly, the spiking identity/uniqueness of each neuron is largely preserved by this approach as shown qualitatively in Figure \ref{fig:reyes_puerta}A compared to Figure 17A1 in \citealp{markram_reconstruction_2015} (gids preserved), and quantitatively in Figure \ref{fig:reyes_puerta}E.
Appearance of previously reported precisely repeating triplet structures is markedly reduced, but could be compensated by moving the network closer to the critical point.
Correlation structure spatially and temporally is found to be significantly reduced in the simplified circuit, even if the reduced criticality of the circuit is compensated.
This could indicate a shortcoming of point-neurons in general, but might be a short-coming of the specific simplification approach and neuron model taken here.
The existence of these benchmarks to systematically address quality of simplification is an important step towards addressing such questions systematically.

\subsection{Outlook}

One objective of simplification is to reduce required compute resources, or target real-time and accelarated execution on GPU-based and neuromorphic platforms \citep{nageswaran2009configurable, fidjeland2010accelerated, bruderle2009establishing, Galluppi2010, Bruderle2011, yavuz_genn:_2016}.
Targeting a simulator such as BRIAN \citep{goodman2008brian}, NEST \citep{gewaltig2007nest} or Nengo \citep{stewart2009python} , which are optimized for point-neuron simulations is desirable, either directly or through PyNN \citep{davison2009pynn}, as these together account for the majority of the point-neuron simulation user community, according to a recent survey \citep{hanke2011neuroscience}.
For these simulators, a common optimization technique is to employ linear synapse models, and lump them all into one state variable.  This approach is not immediately possible with the method proposed here, because the synapses have a variety of time constants determined by their dendritic location, and variability in parameters inherited from the reference model.
Nevertheless we could already show for the single neuron case (Figure \ref{fig:syn_replaced_results}B) that it is possible to reduce the large number of thousands of post-synaptic processes to 3-36 processes per cell with only moderate increase of error by using k-means clustering of the synaptic delay $\tau$.
Future work will further analyse and validate this technique to simplify synapses for running on these popular simulators.

An important opportunity which arises with the availability of data-driven point neuron network is to utilize them in turn as a reference model for the subsequent simplification to population density approaches, for which ample tools and analytical strategies already exist\citep{renart2004mean, muller2007spike}.  Such models are highly suitable for mathematical analysis of population dynamics.

The reference model used here is itself a moving target, and will be refined over time to incorporate new data on gap junctions, interactions with extracellular space, emphatic effects, plasticity, glial cells, and neural-glial-vasculature interactions.  Further simplification techniques may need to be developed to simplify these new aspects to point-neuron representations.  Herein lies the fundamental important of an automated simplification pipeline presented here. Coupled with the data-driven reference model which is continuously evolving to integrate experimental data, a continuous bridge for exchange is achieved between experimental neuroscience and data-driven ``bottom-up'' models on the one side, and the predominantly ``top-down'' point-neuron modeling community on the other.

\section*{Disclosure/Conflict-of-Interest Statement}

The authors declare that the research was conducted in the absence of any commercial or financial relationships that could be construed as a potential conflict of interest.

\section*{Acknowledgments}

\textit{Funding\text{\rm :}} The work was supported by funding from the EPFL to the Laboratory of Neural Microcircuitry (LNMC) and funding from the ETH Domain for the Blue Brain Project (BBP).
Additional support was provided by funding for the Human Brain Project from the European Union Seventh Framework Program (FP7/2007- 2013) under grant agreement no. 604102 (HBP).
The BlueBrain IV BlueGene/Q and Linux cluster used as a development system for this work is financed by ETH Board Funding to the Blue Brain Project as a National Research Infrastructure and hosted at the Swiss National Supercomputing Center (CSCS).

\bibliographystyle{chicago}
\bibliography{paper}

\newpage

\end{document}